\documentclass[aps,prd,11pt,twocolumn,superscriptaddress,notitlepage,nofootinbib,tightenlines,groupedaddress]{revtex4-1}

\usepackage{amsmath,amssymb,amsfonts}
\usepackage{graphicx}
\usepackage{subfigure}
\usepackage{color}
\usepackage{hyperref}
\definecolor{darkred}{rgb}{0.8,0.1,0.1}
\hypersetup{colorlinks=true, linkcolor=darkred, citecolor=blue, linktoc=page}

\newcommand{\N}[1]{\ensuremath{\mathcal N\,{=}\,#1}}

\DeclareMathOperator{\tr}{tr}

\makeatletter\def\l@subsubsection#1#2{}%
\makeatletter\def\l@subsection#1#2{}%
\makeatother

\newcommand{\U}[1]{\ensuremath{\mathsf{U}(#1)}}
\newcommand{\SO}[1]{\ensuremath{\mathsf{SO}(#1)}}

\begin{document}

\title{Holographic entanglement entropy and the internal space}

\author{Andreas Karch}
\email{akarch@uw.edu}
\author{Christoph F.~Uhlemann}
\email{uhlemann@uw.edu}
\affiliation{Department of Physics, University of Washington, Seattle, WA 98195-1560, USA}

\begin{abstract}
We elaborate on the role of extremal surfaces probing the internal space in AdS/CFT.
Extremal surfaces in AdS quantify the ``geometric'' entanglement between different regions in physical space for the dual CFT.
This, however, is just one of many ways to split a given system into subsectors, and extremal surfaces in the internal
space should similarly quantify entanglement between subsectors of the theory.
For the case of AdS$_5\times$S$^5$, their area was interpreted as entanglement entropy between \U{n} and \U{m}
subsectors of \U{n\,{+}\,m} \N{4} SYM.
Making this proposal precise is subtle for a number of reasons, the most obvious being that from the bulk one usually
has access to gauge-invariant quantities only, while a split into subgroups is inherently gauge variant.
We study \N{4} SYM on the Coulomb branch, where some of the issues can be mitigated and the proposal can be sharpened.
Continuing back to the original AdS$_5\times$S$^5$ geometry, we obtain a modified proposal, based on the relation of the
internal space to the R-symmetry group.
\end{abstract}

%%%%%%%%%%%%%%%%%%%%%%%%%%%%%%%%%%%%%%%%%%%%%%%%%%%%%%%%%%%%%%%%%%%%%%%%%%%
\maketitle
\tableofcontents

\section{Introduction}
A fascinating aspect of AdS/CFT is how properties of the CFT are geometrized in the bulk description.
Understanding that relation allows to address questions about the bulk quantum gravity using field-theory methods,
which from a conceptual point of view may be the most interesting application of the dualities.
When it comes to explicitly reconstructing the bulk geometry from the dual field theory,
entanglement correlations play a prominent role \cite{Ryu:2006bv, VanRaamsdonk:2010pw}.
Using the entanglement 1$^\mathrm{st}$~law in the CFT,
it is actually possible to derive the linearized bulk gravity field
equations from the CFT \cite{Lashkari:2013koa,Bhattacharya:2013bna,Faulkner:2013ica,Swingle:2014uza}.
So far, the internal space of the bulk geometry has played a very subordinate role in the relation of
entanglement entropy (EE) to bulk minimal surfaces \cite{Ryu:2006bv, Hubeny:2007xt}, and the same applies
for the procedures to reconstruct the bulk geometry from boundary data.
On the other hand, the internal space is crucial for the dualities already when it comes to matching the symmetries on both sides.

The intriguing proposal of \cite{Mollabashi:2014qfa} aims to identify the area of codimension-2 minimal surfaces wrapping an entire spatial slice
of the AdS factor of AdS$_5\times$S$^5$ with the entanglement entropy between \U{n} and \U{m} subsectors of \U{n\,{+}\,m} \N{4} SYM.
The proposal passed a number of consistency checks, including the behavior of the entropy as function of the ratio $n/m$ and the fact that it
is proportional to the volume of the space on which the field theory is defined.
Nevertheless, there are a number of rather unsatisfactory features, too.
The usual definition of entanglement entropy builds on a tensor decomposition of the Hilbert space,
and one may wonder whether there is a gauge-invariant way to specify the desired subsectors.
On top of that, the degrees of freedom in the two subsectors do not even add up to those of the full theory:
it is not clear how to treat the fields in \U{n\,{+}\,m}/(\U{n}$\otimes$\U{m}), which we will collectively refer to as the ``Ws" in analogy with the W-bosons of the standard model.
Another issue is that extremal surfaces with general Dirichlet boundary conditions at
the boundary of AdS do not even exist \cite{Graham:2014iya}.
Any attempt to directly interpret such a boundary condition as specifying the split into subsectors
therefore seems questionable.

For these reasons we start out from \U{n\,{+}\,m} \N{4} SYM on the Coulomb branch\footnote{This procedure had already been mentioned in \cite{Mollabashi:2014qfa} as a motivation for their proposal.},
where the gauge symmetry is spontaneously broken to \U{n}$\otimes$\U{m}.
When a UV cut-off is imposed far below the mass scale of the Ws, one can actually make the proposal precise. The only degrees of freedom left in the low energy theory are the fields of the unbroken \U{n}$\otimes$\U{m} and one can indeed calculate the entanglement between the two independent gauge sectors from a minimal surface. By raising the UV cut-off one can
gradually add the heavy fields back into the picture.
The bulk geometry is a multi-center brane solution, which is AdS$_5\times$S$^5$ only asymptotically.
We impose boundary conditions on the minimal surface in the IR, where their interpretation can be
understood more straightforwardly than in the UV.
This, however, makes the departure from AdS$_5\times$S$^5$ crucial for understanding the interpretation of the minimal surfaces.
We are led to a sharpened version of the proposal of \cite{Mollabashi:2014qfa}, where the additional fields in
\U{n\,{+}\,m}/(\U{n}$\otimes$\U{m}) play a crucial role.
In fact, we find that the entanglement among these Ws dominates the entanglement entropy when the UV cut-off is large.
As the UV cut-off becomes very large, the minimal area eventually becomes insensitive to the details of the split into
subgroups, indicating that this is not quite the way to look at it.
We propose a new identification of the area of minimal surfaces dividing the internal space with entanglement entropies,
which is based on the global symmetries involved and closer to the usual AdS/CFT dictionary.

\section{Bulk geometry for \texorpdfstring{\N{4}}{N=4} SYM on the Coulomb branch}
To fix notation we introduce the bulk geometry and briefly emphasize some of the properties relevant here.
The metric for the IIB supergravity solution corresponding to two separated stacks of D3 branes can
be written as 
\begin{align}\label{eqn:brane-stacks-metric}
\begin{split}
 ds^2&=f^{-1/2}\eta_{\mu\nu}dx^\mu dx^\nu+f^{1/2}d\vec{y}^{\,2}~,\\
 f&=1+\frac{\kappa R^4}{|\vec{y}-\vec{Y}_1|^4}+\frac{(1-\kappa) R^4}{|\vec{y}-\vec{Y}_2|^4}~.
\end{split}
\end{align}
We set the radius of curvature to $R\,{=}\,1$ in the following, and note that in the usual limits of small string length, large
$N$ and large 't Hooft coupling only the last two terms in $f$ survive. $\vec{Y}_1$ and $\vec{Y}_2$ correspond to the positions of the two
stacks of D3 branes. Without loss of generality we can take their separation to be along the $y_1$ direction and choose the 
origin of the transverse space half way between the brane stacks so that $\vec{Y}_{1,2} = (\pm d,\vec{0})$.  
$\kappa\equiv n/(n+m)$ parametrizes the relative size
of the two stacks. 
The stack at $y_1=+d$ consists of $n$ coincident D3 branes, whereas the stack at $y_1=-d$ consists of $m$ D3 branes.

We parametrize the space transverse to the D3 branes
by setting $y_1\,{=}\,y$ and
$y_i\,{=}\,r\omega_i$ for $i\,{=}\,2, \ldots,6$ with $\sum_i\omega_i^2=1$, such that
\begin{align}
\begin{split}
 d\vec{y}^2&=dy^2+dr^2+r^2d\Omega_4^2~,\\
 f&=\frac{\kappa}{\left((y+d)^2+r^2\right)^2}+\frac{1-\kappa}{{\left((y-d)^2+r^2\right)^2}}\,.
\end{split}
\end{align}
This makes manifest the \SO{5} rotational symmetry in the $y_2, \ldots, y_6$ directions.
A connection to the standard Poincar\'{e} AdS$_5\times$S$^5$ metric can be made by setting
\begin{align}\label{eqn:AdS5S5-coords}
 r&=u\sin\theta& y&=y_m+u\cos\theta~,
\end{align}
where $y_m=d(1-2\kappa)$ is the location of the maximum of $f$ on slices of constant $r$ for large $r$,
and gives the center of mass of the brane stacks \cite{Klebanov:1999tb}.
For $u\,{=}\,|\vec{y}|\,{\gg}\, 1$ we get $f\approx R^4/|\vec{y}|^4$ and this yields the
Poincar\'{e} AdS$_5\times$S$^5$ metric with conformal boundary at $u=\infty$.

The geometric data $d$ and $\kappa$ describing this 2-centered solution has a direct field theory interpretation. 
Due to the $[X_i,X_j]^2$ potential for the 6 adjoint scalars of ${\cal N}=4$ SYM, the moduli space is parametrized 
by 6 commuting matrices in the adjoint of \U{n\,{+}\,m}. Since they are commuting, they can be simultaneously diagonalized. 
The eigenvalues can then directly be interpreted as the $\vec{y}$-space positions of the corresponding D3-branes. 
The 2-centered solution described in here corresponds to a locus where \U{n\,{+}\,m} is broken to \U{n}$\otimes$\U{m} by a 
vacuum expectation value set by $d$. Concretely, the W mass is given by the energy of a string stretched between the two 
stacks, $m_W = \frac{\sqrt{\lambda} d}{\pi}$.

\subsection{Introducing a UV cut-off}
\begin{figure}[ht]
\center
  \includegraphics[width=0.8\linewidth]{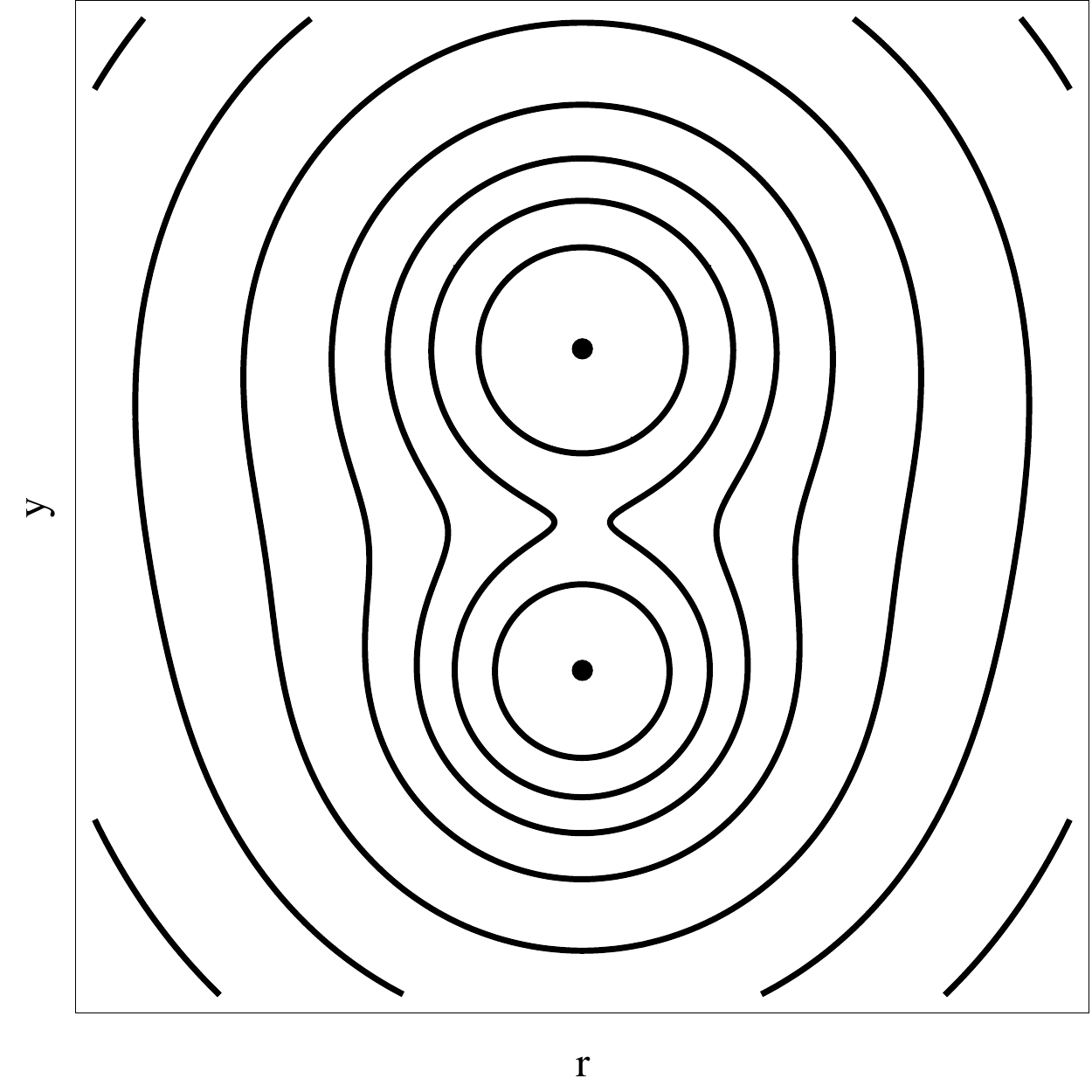}
\caption{Level sets of $f$ for $\kappa=\frac{1}{3}$, giving the cut-off surfaces.
         The two black dots mark the positions of the brane stacks.\label{fig:level-sets-f}}
\end{figure}

Studying the dual CFT with an explicit UV cut-off will be an essential part of what follows.
In the usual AdS/CFT prescription this would translate to a large-volume cut-off in the dual AdS$_5$ geometry, removing the region of space with $u \geq u_*$.
Intuitively speaking, the scale factor multiplying the Minkowski factor of the metric then corresponds to the
cut-off scale in the dual CFT.
Here, we do not quite have AdS$_5\times$S$^5$, so also the cut-off prescription looks a bit different.
Like in the usual AdS$_5$ picture, we define the cut-off surface as a codimension-$1$ surface where the scale factor
multiplying the Minkowski part of the metric (\ref{eqn:brane-stacks-metric})  is constant.
This gives the level sets of $f$ as cut-off surfaces, as shown in Fig.~\ref{fig:level-sets-f}.
The precise relation of the bulk cut-off to a cut-off in the dual CFT is subtle, and not completely
understood \cite{Faulkner:2010jy,Heemskerk:2010hk}.
Here we will only use the qualitative picture, that a larger bulk cut-off (meaning smaller $f$) corresponds to including more UV degrees of freedom in the dual CFT.

\subsection{Connected vs.\ disconnected bulk}
The geometric and topological properties of the bulk geometries with a cut-off at the surface $f=\Lambda^{-4}$ are quite different
for different values of $\Lambda$ (which corresponds to an energy scale in the dual CFT).
For small enough $\Lambda$ we find two disconnected components.
Close to each one of the brane stacks, the influence of the other one is negligible,
and we thus get just two CFTs with gauge groups \U{n} and \U{m}.
This is reflected in the cut-off surface being spherical around each one of the brane stacks.
There's no interactions between them. For operators of large scaling dimension this may be seen from the fact that their
correlators can be computed from bulk geodesics, and there just are no geodesic connecting the two components.

As the cut-off is increased, such that more of the bulk geometry is included around the brane stacks,
we see that the throats start slightly deforming. This is a sign of the fact that the holographic cut-off
is not a strict UV cut-off that completely removes all degrees of freedom above a given energy.
Some parts of the Ws are still present in the cut-off theories. We thus get two deformed CFTs,
which include \N{4} SYM with gauge group \U{n} and \U{m}, respectively, and some parts of the Ws in addition.
There are still no interactions yet between the two subsectors.

As $\Lambda$ is increased further, the bulk geometry eventually becomes connected.
The point where the two components meet is at the minimum of $f$ on the slice $r=0$, which is
\begin{align}\label{eqn:y-star}
y_\star&=\frac{1-x}{1+x}d~,& x&=\left(\kappa^{-1}-1\right)^{1/5}~.
\end{align}
The critical value of $\Lambda$, where the components meet, is $\Lambda=f(y_\star)^{-1/4}\,{\propto}\, d$.
Since $d$ sets the symmetry breaking scale and hence the mass of the massive gauge bosons in $\U{n\,{+}\,m}/(\U{n}\otimes\U{m})$,
we apparently have now crossed their mass scale.
This way we get interactions among the two subsectors.
The fact that there is a sharp transition between interacting and non-interacting CFTs despite the fact
that the Ws are never really cut out completely seems puzzling at first.
But this is precisely the sharp transition found in \cite{Aprile:2014iaa} from a different analysis.
Further increasing $\Lambda$, there is another special value, beyond which the shape of the bulk geometry becomes convex.

As indicated above, the picture we get from just studying the cut-off bulk geometries is reminiscent of
the analysis in \cite{Aprile:2014iaa}, which also analyzed entanglement entropies on the Coulomb branch.
Our analysis in the following sections differs in crucial ways:
we will be using extremal surfaces spanning the whole field theory space and impose an actual UV cut-off.
Instead, \cite{Aprile:2014iaa} used the geometric entanglement entropy for spheres of increasing radius to probe the IR.
In that context the field-theory meaning of the construction is clear and the holographic prescription reduces to the RT (Ryu-Takayanagi) minimal surface.
The reason why their calculation was sensitive to details of the internal space is that the geometry was not globally a product space.
In contrast, we try to give a direct interpretation for RT surfaces in the internal space.

\section{One-parameter family of extremal surfaces separating the brane stacks}

In the standard RT description the entanglement entropy is computed as the area of an extremal surface separating the two entangled regions \cite{Ryu:2006bv}. In the same spirit, entanglement between the degrees of freedom in the two D3 brane stacks should be encoded in extremal surfaces separating the two brane stacks, which we set out to study in this section.
Note that a surface of this class can not be transformed continuously into a surface which does not separate the brane stacks.
This is despite the fact that the area stays finite as one crosses the singularities at $r=0$, $y=\pm d$.

The background geometry has an \SO{5} rotational symmetry in the $y_2,
\ldots, y_6$ directions,
and we will look for minimal surfaces invariant under these symmetries.
The point on the Coulomb branch we are considering preserves this symmetry, and we will look for entanglement entropies
which preserve it as well\footnote{%
In the deep IR, where the split into two subgroups becomes precise, the full \SO{6} R-symmetry is restored.}.
More precisely, we consider a slice of constant time (the setup is static) and look for
extremal surfaces separating the two brane stacks. These can then be parametrized by $y=y(r)$.
The area of such a surface reads
\begin{align}\label{eqn:area}
A&=V^{}_{\mathrm{S}^3}V^{}_4\int_0^\infty dr \,r^4\sqrt{1+y^\prime(r)^2}f^{1/2}~,
\end{align}
where $V_{\mathrm{S}^4}$ denotes the volume of an S$^4$ with unit radius and $V_3$ is the volume in the $\vec{x}$ directions.
The Euler-Lagrange equation for extremality of the surface reads
\begin{align}\label{eqn:Euler-Lagrange-eq}
\begin{split}
 r^4\sqrt{1+{y^\prime}^2}\frac{\delta f}{\delta y}
    -\frac{r^4 y^\prime}{\sqrt{1+{y^\prime}^2}}f^\prime&\\
    -2f\frac{d}{dr}\frac{r^4y^\prime}{\sqrt{1+{y^\prime}^2}}&=0~.
\end{split}
\end{align}
For $\kappa\,{=}\,1/2$ we immediately find the solution $y\equiv 0$.
In the language of the asymptotic AdS$_5\times$S$^5$ geometry, this directly
corresponds to the $\theta\equiv\frac{\pi}{2}$ solution found in \cite{Mollabashi:2014qfa}.

We are interested in solutions separating the brane stacks at $r\,{=}\,0$, $y\,{=}\,\pm d$, so
it is natural to impose boundary conditions at $r\,{=}\,0$.
We would na\"ively expect a two-parameter family of solutions to the second-order
differential equation (\ref{eqn:Euler-Lagrange-eq}), labeled by, e.g., $y(0)$ and $y^{\prime}(0)$.
As often the case in AdS/CFT, we will argue that requiring regularity of the solution at $r=0$ in fact imposes a relation between $y$ and $y'$ at $r=0$ and so regular solutions are uniquely determined by $y(0)$.
We will explicitly show that this is indeed the case for fluctuations around the $y\,{\equiv}\, 0$ solution for $\kappa\,{=}\,1/2$,
before coming to the full non-linear case.

The corresponding asymptotic UV behavior is insensitive to the Coulomb branch deformation and hence follows from the 
analysis \cite{Graham:2014iya} of general extremal surfaces asymptoting to AdS$_{k+1}\times$S$^\ell$ in an asymptotically 
AdS$_5\times$S$^5$ geometry. Unlike extremal surfaces in asymptotically AdS, which can end on any prescribed boundary submanifold, 
the internal part of any extremal manifold has to itself be extremal on the boundary. That is, all our extremal surfaces will 
end at an equatorial S$^4$, which corresponds to $\theta=\pi/2$. 
Our numerical simulations are consistent with this statement, as are the calculations in \cite{Mollabashi:2014qfa}, which had 
to truncate the surface at a finite $u$ in order to have it end at other values of $\theta$. 
Taking $\theta$ to be independent of the internal coordinates to preserve the full rotational symmetry, 
the scaling exponents with which it approaches $\theta=\pi/2$ are complex and the asymptotic behavior becomes
\begin{align}\label{eqn:theta-asmypt}
\begin{split}
 \theta =\pi/2& + a u^{-3/2}\cos\Big(\frac{\sqrt{7}}{2}\log u\Big)
 \\&+b u^{-3/2}\sin\Big(\frac{\sqrt{7}}{2} \log u\Big) + \dots~.
 \end{split}
\end{align}
The regularity constraint at $r=0$ fixes the relation between $a$ and $b$. Different values of $y(0)$ will give rise to different values of $a$ asymptotically.

\subsection{Fluctuations around \texorpdfstring{$y\equiv0$}{y=0} at \texorpdfstring{$\kappa=1/2$}{k=1/2}}
We fix $\kappa\,{=}\,1/2$ and linearize (\ref{eqn:Euler-Lagrange-eq}) around $y\,{\equiv}\, 0$, which yields
the equation for fluctuations around that minimal surface. With $\delta=d^2+r^2$ we get
\begin{align}
 2 r \left(r^2-5 d^2\right) y+ \delta\left(d^2+\delta \right) y'+\delta^2  r y''=0\,.
\end{align}
Solving the indicial equation for $y=r^\gamma\sum \alpha_i r^i$ around $r\,{=}\,0$ yields $\gamma\,{\in}\,\lbrace-3,0\rbrace$,
and a general solution can thus be written as $y=r^{-3}y_a+y_b$.
To get an extremal surface separating the two brane stacks, we need a finite $y(0)$, and thus have to fix $y_a\,{\equiv}\,0$.
This yields a one-parameter family of solutions parametrized by $y(0)=:y_0$, as expected.

\subsection{Numerical solutions for the general case}
For the general solutions to (\ref{eqn:Euler-Lagrange-eq}), the ansatz $y\,{=}\,r^\gamma\sum \alpha_i r^i$
does not lead to a simple indicial equation with an a priori fixed number of solutions.
Nevertheless, fixing $\gamma\,{=}\,0$ leads to a recursive relation fixing $\alpha_i$ for $i\,{>}\,1$ in terms of $\alpha_0$.
The other solution we had seen in the linearized case, $\gamma\,{=}\,{-}\,3$, does not in general yield a solution anymore.
Nevertheless, we still expect a two-parameter family of solutions to the second-order ODE.
Indeed, specifying initial data $\lbrace y(r_0)$, $y^\prime(r_0)\rbrace$ at a generic point $r_0>0$ yields two classes of solutions:
In the generic case the solution diverges towards small $r$ at a finite $0\,{<}\,r_\mathrm{min}\,{<}\,r_0$, and the same happens towards large $r$,
where the solution again diverges at an $\infty\,{>}\,r_\mathrm{max}\,{>}\,r_0$.
On the other hand, by tuning the initial data one can arrange for the solution to stay bounded as $r\rightarrow 0$.
In that case it also stays bounded as $r\rightarrow\infty$, and we recover the one-parameter family of bounded solutions
found before as an expansion around $r=0$.
Since we are interested in solutions with finite $y(0)$, these are the surfaces we are looking for\footnote{%
The other solutions correspond to minimal surfaces starting at a point of the S$^5$ at the boundary of AdS,
from where they blow up and extend into AdS, but not enough to reach the two brane stacks or separate them.
}.

The strategy for finding numerical solutions is as follows.
For a given starting value $y_0$, we solve for the first couple of coefficients in the Taylor expansion analytically.
This yields a decent approximation $\tilde y$ to the corresponding solution in a vicinity of $r\,{=}\,0$.
We then take $\tilde y(\epsilon)$ and $\tilde y^\prime(\epsilon)$ as initial data at an $\epsilon\,{\ll}\, 1$ to numerically solve (\ref{eqn:Euler-Lagrange-eq}).
For $|y_0|\,{<}\,d$ we get an extremal surface separating the brane stacks, as desired.
The AdS$_5\times$S$^5$ surfaces studied explicitly in \cite{Mollabashi:2014qfa}, on the other hand,
correspond to starting values $|y_0|\,{\gg}\,d$, for which the geometry probed by the extremal surface becomes AdS$_5\times$S$^5$.
From their geometric properties it is not quite clear how these surfaces relate to a split into subgroups,
and we will give a different interpretation for their area in Sec.~\ref{sec:min-surfaces-r-symmetry}.

\section{Extremal surfaces and entanglement entropy for interacting subsectors}

We have seen in the previous section that we get a one-parameter family of extremal surfaces separating the brane stacks, i.e.\ the ones with $|y_0|<d$.
The proposal of \cite{Mollabashi:2014qfa} is that the minimal among those computes the EE for two interacting
subsectors of the full dual \U{n\,{+}\,m} SYM.
The proposal was to define
the subsectors as the SYM based on the \U{n} and \U{m} subgroups respectively.
For $\cal{N}$=4 SYM on the Coulomb branch,
with the gauge symmetry spontaneously broken to \U{n}$\otimes$\U{m},
this split makes sense in the IR, that is below the mass of the Ws. For these energies only the degrees of freedom belonging to these two subsectors survive and so it makes sense to separate the remaining degrees of freedom according to which subgroup they belong to. In order to understand this regime, we start with an investigation with a rather low explicit UV cut-off.

\subsection{Low cut-off: two (non-)interacting CFTs}\label{sec:low-cut-off}
\begin{figure}[ht]
\center
  \includegraphics[width=.8\linewidth]{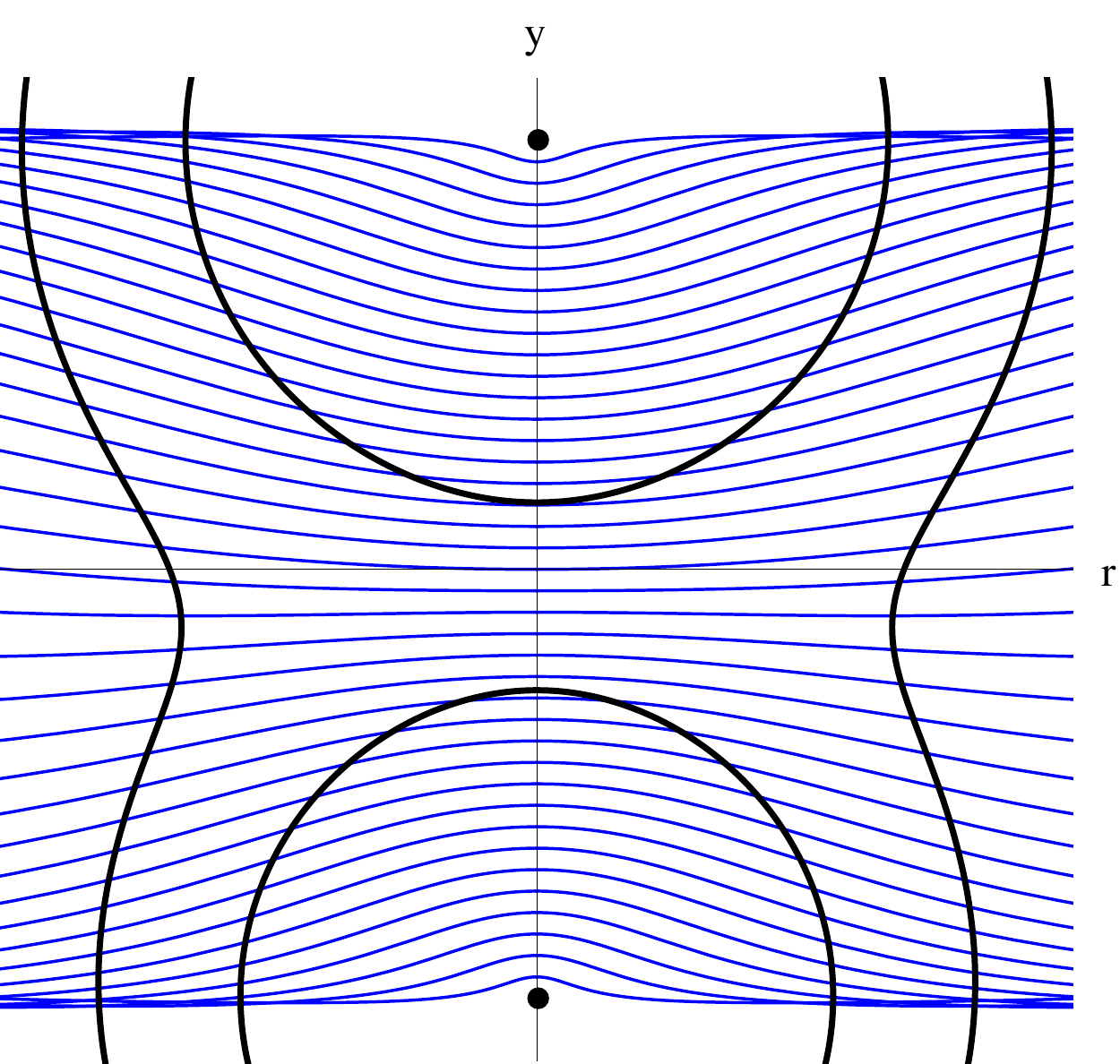}
\caption{One-parameter family of extremal surfaces for $\kappa=1/3$ and we set $d=1$ in all figures.
The cut-off surfaces are shown for $\Lambda^{4}\in\lbrace\frac{2}{3},3\rbrace$.\label{fig:low-cut-off1}}
\end{figure}

We start with the case where the cut-off is below the mass scale of the massive gauge bosons, $\Lambda<f(y_\star)^{-1/4}$.
In that case the bulk geometry has two disconnected components and the subsectors are not entangled.
This is reflected in the properties of the family of extremal surfaces, too.
Fig.~\ref{fig:low-cut-off1} shows the extremal surfaces and two cut-off surfaces, and the case of interest now
corresponds to the inner cut-off surface.
Clearly, the extremal surface with minimal area is one of those starting and ending directly at the cut-off surface, so the EE is zero.
We may still ask what the meaning of the other extremal surfaces separating the branes is.
If the cut-off were a hard UV cut-off in the field theory, the heavy gauge bosons would not play any role whatsoever in this regime, since they are simply cut off.
This is exactly what happens at very low cut-offs, where the bulk geometry to a good approximation consists of just two disconnected cut-off AdS$_5\times$S$^5$ geometries,
and the extremal surface in either one of the components, say the second, define a split where one subsector consists
of CFT$_1$ and part of CFT$_2$, while the other consequently consists of only a part of CFT$_2$.
The associated EE therefore is merely due to an ``unfortunate'' split into subsectors,
in the sense that it does not reflect the EE between the \U{n} SYM and the \U{m} SYM alone.

\begin{figure}[htb]
\center
  \includegraphics[width=0.8\linewidth]{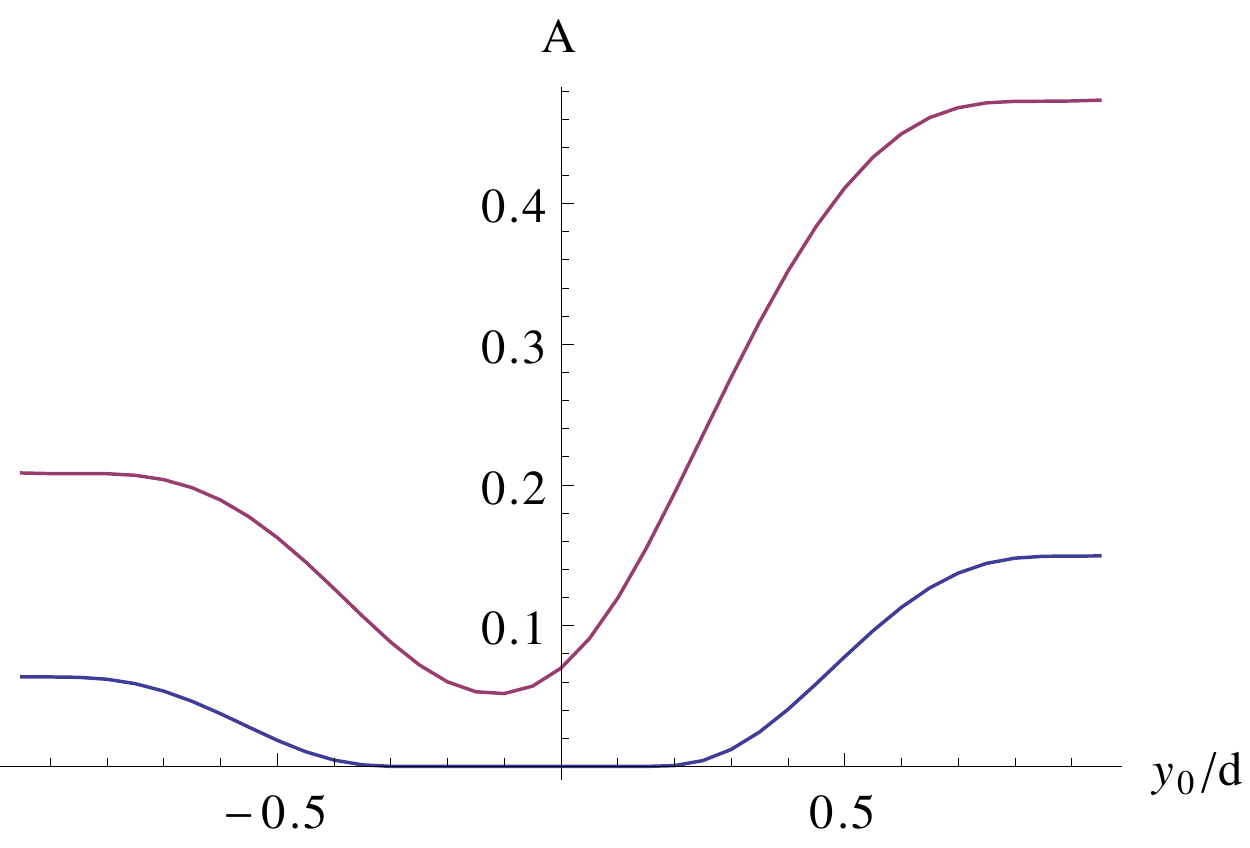}
\caption{Areas of the extremal surfaces as function of the starting value at $r=0$, $y_0$, for the two cut-off
surfaces of Fig.~\ref{fig:low-cut-off1} with the lower/upper curve corresponding to $\Lambda^{4}\in\lbrace\frac{2}{3},3\rbrace$.
\label{fig:low-cut-off2}}
\end{figure}

We now turn to the case where the cut-off is above the mass scale of the heavy gauge bosons,
but still of the same order of magnitude.
The two CFTs now interact non-trivially, as the Ws are part of the spectrum. 
This case corresponds to the outer cut-off surface in Fig.~\ref{fig:low-cut-off1}.
The bulk geometry is still squeezed in the region between the branes.
The area for the family of minimal surfaces is given by the upper curve in Fig.~\ref{fig:low-cut-off2}.
The one with minimal area is one of those very close to the bottleneck, as one would expect intuitively:
The function $f$ has a minimum in between the two stacks and this is where the geometry is very narrow.
The surfaces at the bottleneck therefore have the shortest length in terms of their parametrization
(this is what we see in the picture), but also the proper length per parameter length is minimal.
As $\Lambda$ approaches $f(y_\star)^{-1/4}$, the surface becomes the one starting at $y_\star$ given in (\ref{eqn:y-star}).
For $\kappa=1/2$, the corresponding minimal surface would be the $y\equiv 0$ or $\theta\equiv\pi/2$
surface discussed in \cite{Mollabashi:2014qfa}.

Together with the previous discussion of the disconnected case, we can now give a suggestive argument for why we choose the minimal
among the extremal surfaces to compute the EE.
The difference is not as qualitative anymore, since in this case we can not cleanly associate a single component of the bulk
geometry to a single subsector.
In the field theory this corresponds to the presence of the massive gauge bosons.
Not only do they mediate interactions between the \U{n} SYM and \U{m} SYM, we also
have to decide how to split them and assign them to the two subsectors. For these cut-offs of order the W mass, the EE is still dominated by the entanglement between the two unbroken subgroups, but we can continuously change the EE by shifting how to split the Ws between the two subsectors.
This is what is accomplished by changing $y_0$.
We are interested in the minimal EE that can be achieved,
which we may call {\it irreducible EE}. It is this quantity that we want to continue to identify as the EE between the two unbroken subgroups due to the interactions mediated by the Ws.
This is given by the area of the minimal among the extremal surfaces separating the branes.
Fig.~\ref{fig:low-cut-off2} shows that this is positive.

\begin{figure}[ht]
\center
  \includegraphics[width=0.8\linewidth]{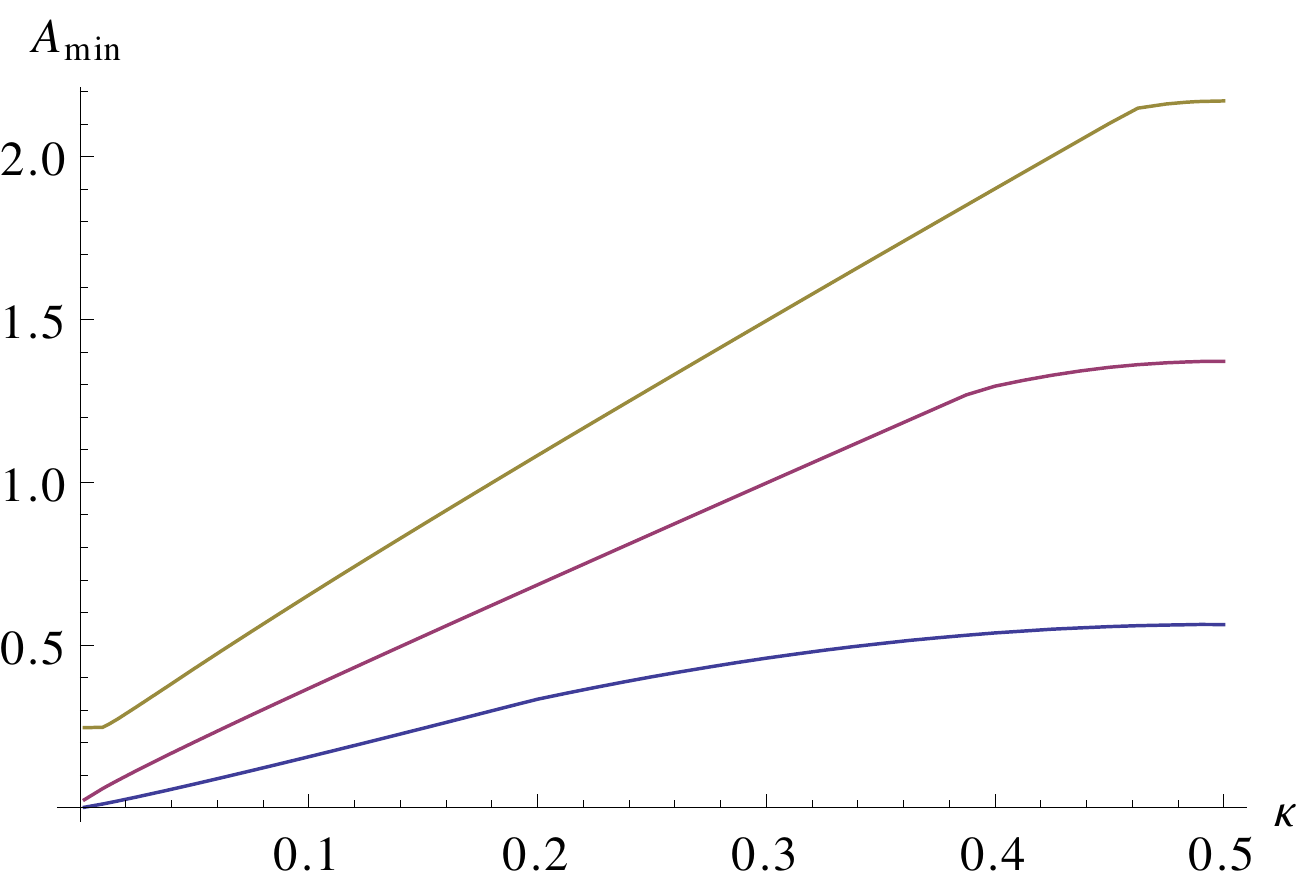}
\caption{Areas of the minimal among the extremal surfaces as function of $\kappa$.
         From lower to upper curve, $\Lambda^{4}\in\lbrace 10,20,30\rbrace$.
         The plot is symmetric in $\kappa\rightarrow 1-\kappa$.
         As $\kappa\rightarrow 0$, we see that the area does not vanish.
        \label{fig:EE-kappa}
         }
\end{figure}

We close the subsection with a look at the EE as a function of $\kappa$.
The result is shown in Fig.~\ref{fig:EE-kappa}. %
One should be suspicious whether it makes sense to compare the computations for different $\kappa$ at a fixed cut-off $\Lambda$. The relation of bulk to boundary cut-off is subtle, making it hard to ensure that the true field theory cut-off stays fixed as we change $\kappa$. In here we simply proceed with the comparison at fixed $\Lambda$, hoping that this will at least give qualitatively correct results.
The result is shown in Fig.~\ref{fig:EE-kappa}.
If we make a crude approximation, assuming that each degree of freedom in one sector is to some extent entangled
with each d.o.f.\ in the other, we should get $S_\mathrm{EE}\propto\kappa^2(1-\kappa)^2$.
This suggests that the EE should be maximal for $\kappa=\frac{1}{2}$ and minimal for $\kappa\in\lbrace0,1\rbrace$.
The results roughly reproduce this anticipated behavior.
They are also roughly compatible with the results found in \cite{Mollabashi:2014qfa}, although obtained in a completely
different way.
Note that the EE does not vanish as $\kappa\rightarrow 0/1$.
Geometrically, this is easily understood from the fact that,
as long as $\Lambda>f(y_\star)^{-1/4}$, the cut-off surface does not come arbitrarily close to the brane stacks, such
that each extremal surface separating the branes necessarily has some finite area.

\subsection{Raising the UV cut-off: the \texorpdfstring{1$^\mathrm{st}$-order}{first-order} phase transition}
\label{sec:first-order-phase-transition}

Our analysis of the low cut-off configurations helped us to understand two important lessons. First, the role of $y_0$ is to determine how the Ws are split between the two subsectors. Second, the corresponding ambiguity in the EE can be uniquely fixed by singling out the minimal EE for a given point on the Coulomb branch and a given cut-off. With these lessons in mind, we now study the behavior as the cut-off is increased further.
The first thing to notice is that the shape of the bulk geometry becomes convex as the cut-off is increased
beyond a certain value, as seen in Fig.~\ref{fig:level-sets-f}.
A second thing to notice is that the structure of $f$ as function of $y$ changes as we move away from the brane stacks.
At $r=0$, $f$ diverges/is maximal at the brane stacks and has a minimum in between.
This behavior  persists to other slices of constant $r$ close to the branes.
Asymptotically, however, the geometry and $f$ approach the behavior for one stack of branes, i.e.\ AdS$_5\times$S$^5$.
This means that $f$ just has a maximum somewhere between $y=d$ and $y=-d$, and no minima.
The surface which starts out as a minimal surface in the vicinity of $r=0$ thus does not necessarily minimize the area for large $r$.
This is most clear for $\kappa=1/2$: close to $r=0$, $y\equiv 0$ is the minimal among the extremal surfaces.
At $r\rightarrow\infty$, however, it sits right on the maximum of $f$, and thus picks up larger contributions than the other extremal surfaces.

\begin{figure}[htb]
\center
\subfigure[][]{ \label{fig:phase-transition1}
  \includegraphics[width=0.8\linewidth]{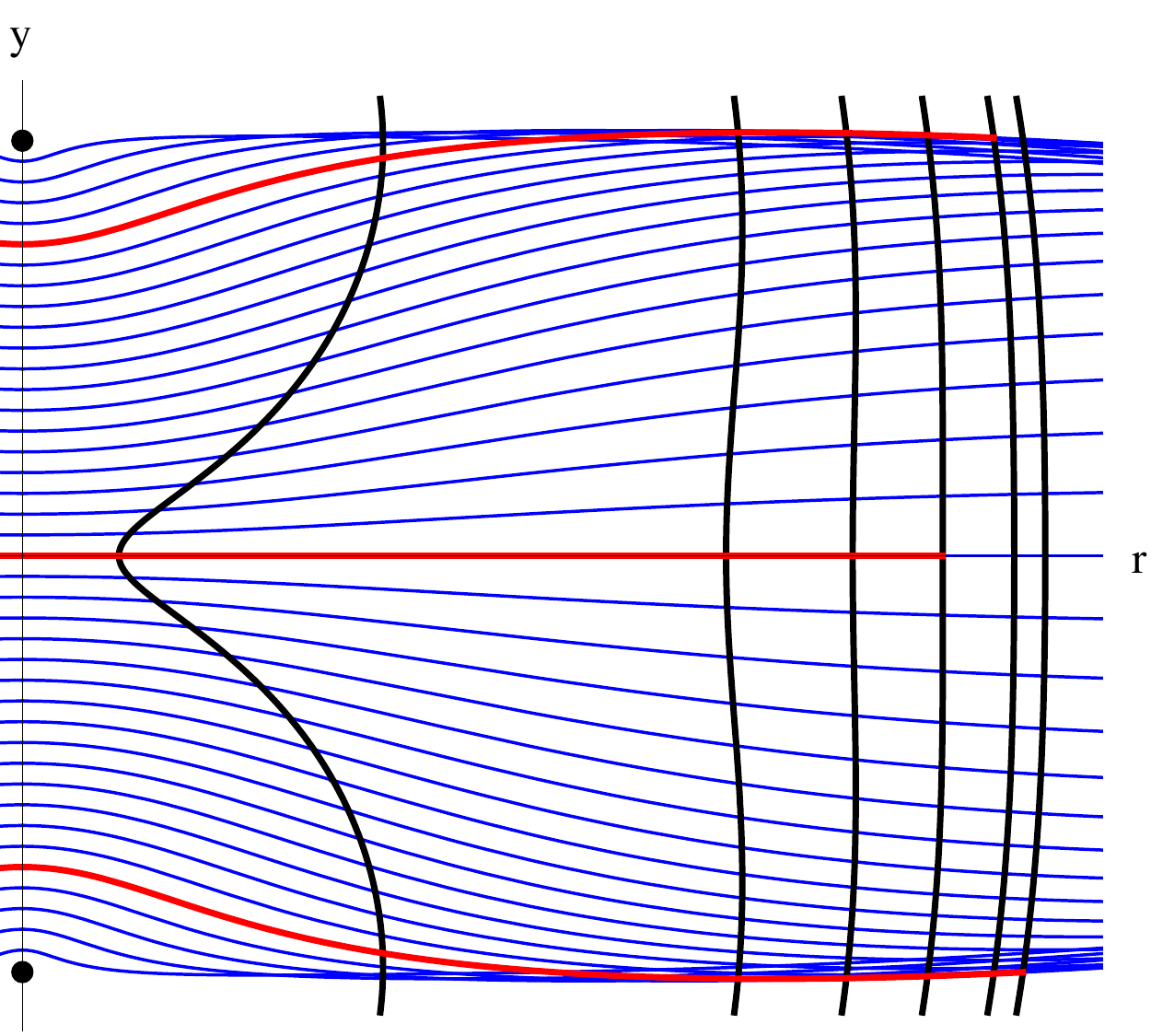}
} \\
\subfigure[][]{ \label{fig:phase-transition2}
  \includegraphics[width=0.8\linewidth]{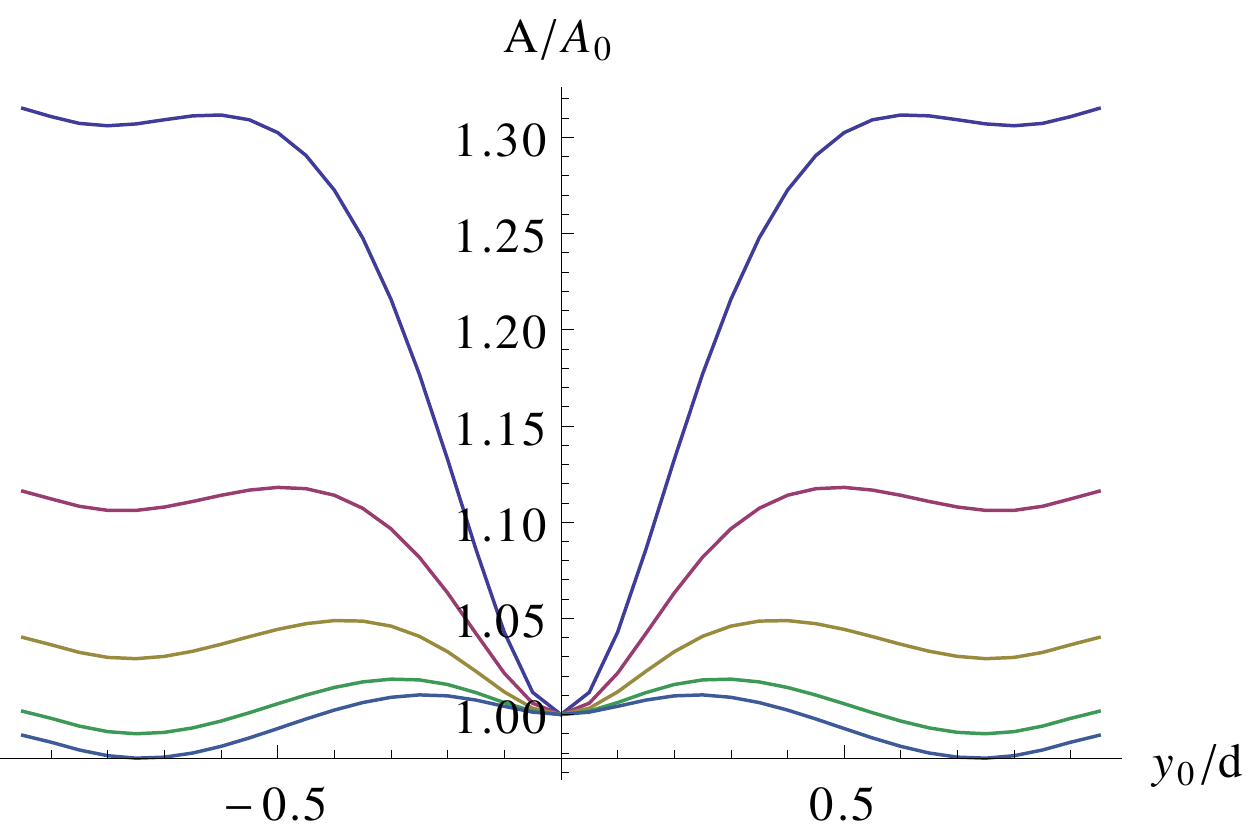}
}
\caption{The top panel shows the one-parameter family of extremal surfaces for $\kappa=1/2$ and
cut-off surfaces for $\Lambda^{4}\in\lbrace 10/9, 15, 25, 35, 45, 50\rbrace$.
The red ones are those with minimal area for a given cut-off (they are red only up to this cut-off).
The bottom panel shows the corresponding areas as function of $y_0/d$.
Each curve is normalized to the area of the $y\equiv 0$ surface
with the corresponding value of the cut-off, and the lower the curve the larger the value of $\Lambda$.
We see that the area develops three local minima, and at a certain value of the cut-off
the two degenerate minima close to the brane stacks become lower than the central one.
This is where the red surface in the upper panel jumps from the center to one of the brane stacks.
\label{fig:phase-transition}}
\end{figure}

This makes us expect a phase transition for some value of the cut-off, where the minimal among the extremal surfaces jumps
from one starting close to $y_\star$ to one starting close to one of the brane stacks.
That this is indeed the case is shown in Fig.~\ref{fig:phase-transition}.
As the cut-off is increased, we see a discontinuous transition for the minimal surface from the surface $y\equiv 0$
to one of the degenerate two starting close to either one of the brane stacks.

With a cut-off only slightly above the mass scale of the heavy gauge bosons,
their sole effect was to mediate interactions between the \U{n} and \U{m} subsectors
and we were able to ignore interactions and entanglement among them.
But with the higher cut-off we get a significant contribution to the EE from how the Ws
are distributed among the subsectors.
They correspond to open strings stretching between the brane stacks, so choosing a minimal surface starting close to one
of the brane stacks seems to correspond to assigning the bifundamentals entirely to one of the subsectors.
The fact that, beyond a certain cut-off, the EE is minimized by the surfaces starting close to either one of the
brane stacks, seems to tell us that the EE is dominated by entanglement among the heavy gauge bosons.
For lower cut-offs it was preferable to start roughly in the middle between the two brane stacks, corresponding to
the cleanest split between the \U{n} adjoint and \U{m} adjoint d.o.f..
But this is outweighed now by the strong entanglement among the heavy gauge bosons,
which means we get the least EE by assigning them to one subsector completely.

\begin{figure}[htb]
\center
\subfigure[][]{ \label{fig:phase-transition3}
  \includegraphics[width=0.8\linewidth]{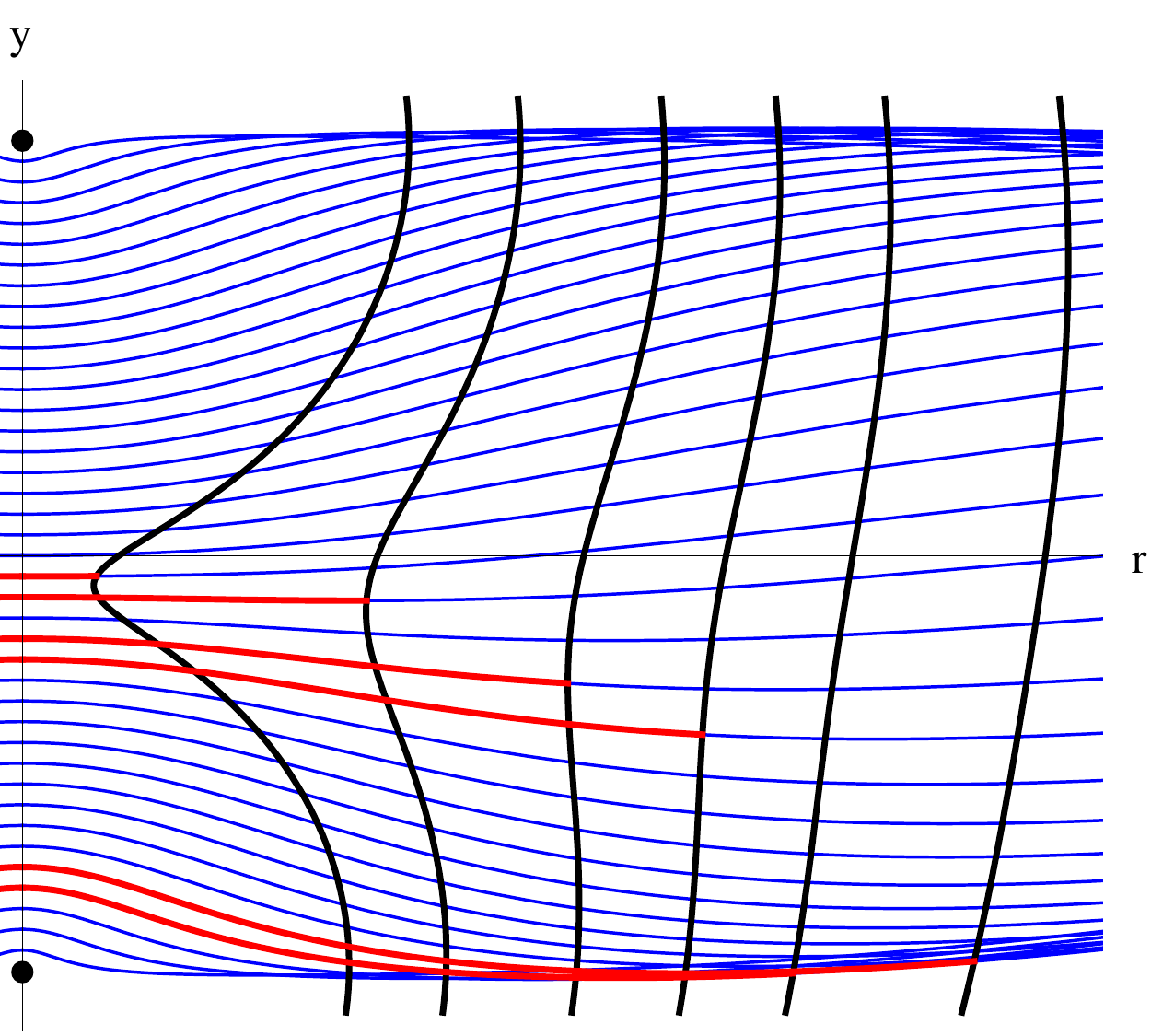}
}\\
\subfigure[][]{ \label{fig:phase-transition4}
  \includegraphics[width=0.8\linewidth]{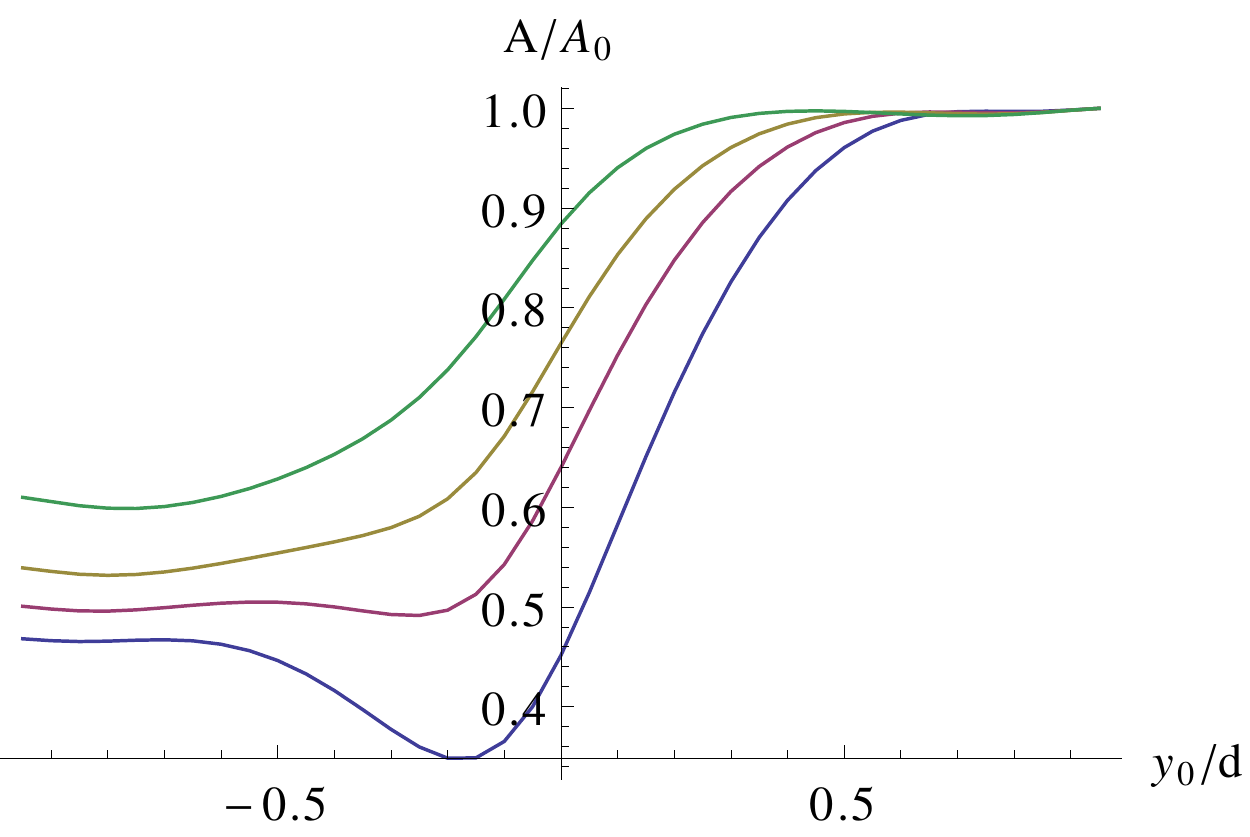}
}
\caption{The top panel shows the one-parameter family of extremal surfaces for $\kappa=1/3$ and
cut-off surfaces for $\Lambda^4\in\lbrace 10/9, 3, 8, 15, 25, 50\rbrace$.
The red ones are again those with minimal area for one of the cut-offs.
The bottom panel shows the corresponding areas as function of $y_0/d$,
with each curve normalized to the area of the surface closest to the brane stack at $y\,{=}\,1$.
\label{fig:phase-transition-asymmetric}}
\end{figure}

Fig.~\ref{fig:phase-transition-asymmetric} shows the same plot for $\kappa=1/3$, to show how the degeneracy
between the two minima is lifted. The qualitative behavior stays the same:
With increasing cut-off the minimal surface slips slightly towards the lighter brane stack, before discontinuously jumping to another one close to the lighter brane stack at a
certain value of the cut-off.
As the cut-off is then increased further, the minimal surface smoothly moves further towards the lighter stack.

To conclude our main results, we have determined the irreducible EE as a function of UV cut-off for a family of Coulomb branch 
configurations where the unbroken gauge group has two factors. 
We found a phase transition that separates the qualitatively different low cut-off and high cut-off regimes. 
With a low cut-off, the entanglement among the Ws can be neglected and the degrees of freedom can be meaningfully split 
according to which subgroup they belong to, along the lines of \cite{Mollabashi:2014qfa}. 
Correspondingly the extremal surface giving rise to the irreducible EE cuts space roughly in the middle between the 
two stacks of D3 branes. At large cut-off, however, the EE is dominated by the Ws. 
The irreducible EE arises for minimal surfaces very close to one or the other stack, 
so that the Ws are almost entirely associated to one or the other subsector. 
Clearly, in this case we can no longer simply tag the two subregions by what unbroken subgroup they belong to.
This becomes even more severe when we look at the UV structure. Since all extremal surfaces dividing the internal space 
end on the same equatorial S$^4$ in S$^5$, they all share the same leading UV divergence in their area.
Clearly, if the areas were intrinsically related to the entanglement entropy between different subgroups, this should not be the case:
different splits would produce different numbers of degrees of freedom in each subsector, which should be reflected in the 
UV structure of the entanglement entropy.
We conclude that the subsectors should rather be defined more directly according to what part in the transverse space they are dual to,
which loosely speaking corresponds to the R-charge. We attempt to make this last statement more precise in the next section.

\section{Extremal surfaces and R-symmetry: a refined proposal}\label{sec:min-surfaces-r-symmetry}
From our explicit investigation of minimal surfaces splitting the internal space in the previous sections,
we have seen that the interpretation of their area as entanglement entropy between \U{n} and \U{m} subsectors of \U{n\,{+}\,m} \N{4} SYM makes sense only in certain regimes.
Moreover, this interpretation is somewhat questionable on formal grounds: from the bulk one usually only has access to gauge-invariant quantities
on the boundary. So a split according to an actual (global) symmetry group, rather than a gauge redundancy group, seems more natural.
With the isometries of S$^5$ corresponding to the R-symmetry group of \N{4} SYM the latter is a natural candidate,
and we elaborate in the following on how such a split could work.

The relation of geometric EE to minimal surfaces in AdS is facilitated by the direct identification of the 
boundary geometry of AdS with the field theory geometry. 
For the internal space this certainly is a bit more tricky, but from the bulk perspective 
CFT subsectors can be assigned to subregions in a qualitatively similar way.
To illustrate that, we start by looking at the geometric EE in the language of
algebraic QFT~\cite{Haag:1992hx}.

The standard definition of EE starts out from a tensor decomposition of the Hilbert space
$\mathcal H=\mathcal H_A\otimes\mathcal H_B$.
The global state is described by a density operator $\rho$, and the reduced density operator for, say, subsystem $A$ is obtained by a partial trace operation $\rho_A=\tr_B\rho$.
The focus in algebraic QFT is more on the algebra of operators and observables, rather than on a concrete Hilbert-space representation.
More precisely, the basic object is a net of operator algebras $O\mapsto \mathcal A(O)$, associating to each region of spacetime $O$
an algebra $\mathcal A(O)$, which is a subalgebra of a $\star$-algebra $\mathcal A$ (e.g.\ an abstract C$^\star$-algebra or a von Neumann algebra).
The self-adjoint elements represent the physical observables, and a
state is described as a map from the algebra to the complex numbers, $\omega:\mathcal A \rightarrow \mathbb{C}$.
In a Hilbert space representation that state can be represented by a density operator $\rho$ via $\omega:\mathcal O\mapsto\tr(\rho\mathcal O)$ for $\mathcal O\in\mathcal A$.
A subsystem corresponds to a subalgebra of $\mathcal A$, and the reduced density operator is the representation of the pullback of the global state to that subalgebra.
A state is pure iff it can not be written as a convex combination of other states,
i.e.\ as $\omega=\alpha\omega_1+(1-\alpha)\omega_2$ with $\alpha\in(0,1)$.
Now it may be possible to find $\omega_{1/2}$ and $\alpha$ to satisfy this equation on a subalgebra, but not on the entire algebra, and this is how a pure state becomes mixed upon restriction.

The geometric EE in AdS/CFT fits into that framework as follows:
Setting the boundary values of bulk fields to zero outside of a region $A$ on some constant-time slice, we only source operators localized in $A$
at that time. 
To define the subsystem associated to $A$, we then take the subalgebra of $\mathcal A$ generated by that 
set of operators, e.g.\ its double commutant in $\mathcal A$ for von Neumann algebras.
The pullback of the global state to that subalgebra via the inclusion map $\iota$ gives the reduced density operator 
$\rho_A\leftrightarrow \iota_\star\omega$,
and the RT proposal \cite{Ryu:2006bv} states that a minimal surface in AdS computes the von Neumann entropy of that pullback state.

Let us now turn to an extension of this proposal to minimal surfaces in the internal space.
The first step is to make sense of what it means to restrict sources to a subspace of the internal space.
To this end we look at boundary data $\phi_0(x,y)$ for a bulk field $\phi(x,z,y)$,
where $x,z$ label coordinates on AdS and $y$ are coordinates on the internal space.
That boundary data may be expanded in spherical harmonics as
\begin{align}
 \phi_0(x,y)&=\sum_{r,\vec{m}} \phi_{0,r,\vec{m}}(x)Y_{r,\vec{m}}(y)~.
\end{align}
The $Y_r$ are spherical harmonics on S$^5$, $r$ runs through the representations of $\SO{6}$ and $\vec{m}$ are the 
analogs of the angular-momentum quantum number.
Each of the $\phi_{0,r,\vec{m}}$ is now identified as source for an operator $\mathcal O_{r,\vec{m}}$ in \N{4} SYM.
Restricting the boundary data to have support only in a part $A$ of S$^5$ (at some given time) corresponds to sourcing only very particular linear combinations of operators.
Namely, we can only source operators
\begin{align}
\mathcal O_A=\sum_{r,\vec{m}} c_{r,\vec{m}}\mathcal O_{r,\vec{m}}~,
\end{align}
where the coefficients $c_{r,\vec{m}}$, if interpreted as coefficients for the spherical harmonics, produce a function with support in $A$ only.
Let us denote this set of operators, which can be sourced by bulk fields which are non-zero only in the part $A$ of S$^5$, by $\mathsf{Op}(A)$.
An extension of the minimal-area prescription to minimal surfaces in the internal space emerges naturally now:
To define a subsystem we take the subalgebra $\mathcal A_A$ of $\mathcal A$ that is generated by $\mathsf{Op}(A)$.
A reduced density operator on this subsector of the theory can again be defined via the pullback of the global state to the subalgebra.
In analogy to the geometric EE, a minimal surface which splits the internal space into $A$ and its complement should then compute the 
von Neumann entropy of this reduced state, giving the desired extension of the RT proposal to the internal space.
The construction can be extended to other choices of the compact manifold or to geometries which are AdS$\times$compact 
only asymptotically by following the same logic.\footnote{%
We have not shown that this yields a tensor decomposition.
So while we have defined an entropy, it is not clear that this is an entanglement entropy in the usual sense.
This plagues geometric EE in gauge theories as well \cite{Casini:2013rba}.
}

We can now rephrase the analysis of the previous sections in the following way:
We studied particular subalgebras of operators, corresponding to the part of the internal space that the minimal surfaces
studied there wrap at the boundary of the cut-off bulk geometry. In principle, they are characterized by their $R$-symmetry representations as just outlined.
By imposing a selection criterion in the IR, we selected the split which in the IR coincides with a split into unbroken sub-gauge-groups.
This corresponded to a particular orientation of the S$^4$ entangling surface w.r.t.\ the alignment of the vev in a Higgsed \N{4} SYM, and we can thus
incorporate the proposal of \cite{Mollabashi:2014qfa} into a more natural picture based on $R$-symmetry.

On the conformal boundary itself, we know already from \cite{Graham:2014iya} that all minimal surfaces splitting the internal space end on
extremal boundary surfaces. This seems to have a natural interpretation in the context of the current construction:
going to the UV, the completion of $\mathsf{Op}(A)$ to an algebra seems to need all the operators corresponding at least to a half sphere, no matter what the region we started.
We leave a more detailed analysis of this issue for the future.

Summing up, we have given a field-theory construction to select a set of operators corresponding to keeping only a part of the internal space in AdS$_5\times$S$^5$.
This set can be completed to a minimal algebra by a unique construction, which allows to associate a well-defined subsystem to them.
The entropy of the corresponding reduced density operator would -- by straightforward extension of the RT proposal -- be expected to be given by the area of a minimal surface splitting the internal space.
This identification seems better motivated than the proposal of \cite{Mollabashi:2014qfa} on formal grounds, 
since in contrast to the gauge group the global R-symmetry group is indeed directly accessible from the bulk.
We found that in the IR the proposal coincides -- for certain cases and up to subtleties we discussed -- with the split
according to subgroups. In the UV, however, it is clearly distinct.

\vfil

\section*{Acknowledgments}
We thank Andy O'Bannon and Alexander Schenkel for helpful discussions, and Tadashi Takayanagi for useful correspondence on \cite{Mollabashi:2014qfa}.
The work of AK was supported, in part, by the US Department of Energy under grant number DE-SC0011637.
CFU is supported by {\it Deutsche Forschungsgemeinschaft} through a research fellowship.

\bibliography{minimal-internal}
\end{document}